\documentclass{aa}
\usepackage{graphicx}
\usepackage{txfonts}
\usepackage{natbib}
\usepackage{tabularx,colortbl,threeparttable}
\usepackage[table]{xcolor}

\bibpunct{(}{)}{;}{a}{}{,} 

\def\msun{{\it M}$_{\sun}$} 

  \definecolor{tableShade}{HTML}{E0E5E9}

\begin{document}

   \title{Rubidium, zirconium, and lithium production in intermediate-mass asymptotic
   giant branch stars}

   \author{M. A. van Raai
          \inst{1}
          \and
          M. Lugaro
          \inst{2}
          \and
          A.I. Karakas
          \inst{3}
          \and
          D.A. Garc\'ia-Hern\'andez
          \inst{4,5}
	  \and
	  D. Yong
	  \inst{3}
          }

   \institute{Sterrekundig Instituut, University of Utrecht,
              Postbus 80000 3508 TA Utrecht, The Netherlands,
              \email{markvanraai@gmail.com}
              \and
             Monash Centre for Astrophysics (MoCA),
            School of Mathematical Sciences, Monash University,
          Clayton Victoria 3800, Australia,
             \email{maria.lugaro@monash.edu}
         \and
	       Research School of Astronomy and Astrophysics, Mt. Stromlo
	       Observatory, Cotter Rd., Weston, ACT 2611 Australia,
             \email{akarakas@mso.anu.edu.au, yong@mso.anu.edu.au}
         \and
	      Instituto de Astrofisica de Canarias, C/ Via L\'actea s/n,
               38200 La Laguna (Tenerife),
             Spain, \email{agarcia@iac.es}
         \and
             Departamento de Astrof\'isica, Universidad de La Laguna (ULL),
              E-38205 La Laguna, Spain
	 }

   \date{Received / Accepted }

\authorrunning{van Raai et al.}
\titlerunning{Rb, Zr, and Li production in intermediate-mass AGB stars}

 \abstract{A recent survey of a large sample of Galactic
 intermediate-mass ($>$3 \msun) asymptotic giant branch (AGB) 
 stars shows that they exhibit large overabundances of rubidium (Rb) 
 up to 100--1000 times solar. 
 In contrast, zirconium (Zr) is not enriched compared to the solar
 abundances in these stars.  These observations set constraints on
 our theoretical notion of the $slow$ neutron capture process ({\it s}
 process) that occurs inside intermediate-mass AGB stars.
 Lithium (Li) abundances are also reported for these stars. 
 In intermediate-mass AGB stars, Li can be produced by proton captures
 occuring at the base of the convective envelope. For this reason the
 observations of Rb, Zr, and Li set complementary constraints on
 different processes occurring in the same stars.}
 {We present predictions for the abundances of Rb, Zr, and Li 
 as computed for the first time simultaneously
 in intermediate-mass AGB star models and compare them to the current observational constraints.}
 {We calculate the Rb, Zr, and Li surface abundances for stellar models
 with masses between 3 and 6.5 \msun\ and metallicities between 0.02 and 0.004.}
 {We find that the Rb abundance increases with increasing stellar mass,
 as is inferred from observations but we are unable to match the
 highest observed [Rb/Fe] abundances. Variations of the reaction rates of the
 neutron-capture cross sections involved with Rb production and the
 rate of the $^{22}$Ne($\alpha,n$)$^{25}$Mg reaction, responsible
 for neutron production inside these stars, yields only modest variations in
 the surface Rb content of $\approx 0.3$~dex. Inclusion of a partial mixing zone (PMZ) to
 activate the $^{13}$C($\alpha,n$)$^{16}$O reaction as an additional neutron
 source yields significant enhancements in the Rb abundance. However
 this leads to Zr abundances that exceed the upper
 limits of the current observational constraints. If the third dredge-up (TDU)
 efficiency remains as high during the final stages of AGB evolution
 as during the earlier stages, we
 can match the lowest values of the observed Rb abundance range. 
 We predict large variations in the Li abundance, which are observed.
 Finally, the predicted Rb production increases with decreasing
 metallicity, in qualitative agreement with observations of Magellanic
 Cloud AGB stars. However stellar models of $Z=0.008$ and $Z=0.004$
 intermediate-mass AGB stars do not produce enough Rb to match the observed abundances.}
 {}



\maketitle

\section{Introduction}

The elements heavier than iron are produced almost entirely by
neutron-capture nucleosynthesis via the $s$(low) and the 
$r$(apid) processes, which are each responsible for roughly half 
of the cosmic abundances of these elements \citep{kaeppeler89}. During
the $s$ process timescales for neutron captures on unstable isotopes 
are typically lower than the decay timescales, which implies neutron 
densities $N_n \sim 10^{7}$ cm$^{-3}$. Conversely, during the 
$r$ process timescales for neutron captures on unstable isotopes 
are higher than decay timescales, which implies $N_n > 10^{20}$ 
cm$^{-3}$. The element rubidium (Rb) is an example of an element 
that may have a significant $r$-process component, where 
$\sim$20-80\% of its solar abundance has been ascribed to the 
$r$ process. The exact value of this component depends on 
the $s$-process model used to calculate it, which can be parametric 
\citep{arlandini99,simmerer04}, stellar \citep{arlandini99}, or a 
chemical evolution model of the Galaxy \citep{travaglio04}. The 
$r$ process contribution also depends on the uncertainties in 
the solar abundances and nuclear physics inputs \citep{goriely99}. 
Zirconium (Zr) on the other hand is a typical element made by the 
$s$ process, with $\sim$15-30\% of its solar abundance ascribed 
to the $r$ process. The $r$ process is believed to occur in 
supernovae and/or neutron star mergers \citep{meyer94}, while 
the $s$-process elements from Sr to Pb are produced in asymptotic 
giant branch (AGB) stars \citep{gallino98}.

The AGB is the final nuclear burning stage in the evolution of stars 
with initial masses in the range $\approx$ 0.8 to 8
\msun\footnote{Hereafter we refer to AGB stars with initial masses 
$<$ 3 \msun\ as ``low-mass'' AGB stars, and to AGB 
stars with initial masses $>$ 3 \msun\ as ``intermediate-mass'' 
AGB stars.}. During the AGB the H-burning shell is responsible 
for the nuclear energy production most of the time, but every 
$10^{3-5}$ years or so the He shell ignites, driving convection 
throughout most of the He intershell, the He-rich region located 
between the H- and the He-burning shells. This is known as a 
thermal pulse (TP).
After a TP has occurred, at which time 
the H-burning shell lies dormant, the convective envelope 
extends inward and may reach the He intershell. This is known 
as the third dredge-up (TDU) and can occur periodically after 
each TP. The TDU mixes nuclear burning products from the He 
and H-burning shells to the stellar surface, increasing the 
abundance of carbon and in some cases leading to C$>$O at the 
stellar surface \citep{herwig05,karakas07b,lugaro11}.

In the He intershell of AGB stars, neutrons for the $s$ process can 
be released by the $^{13}$C($\alpha,n$)$^{16}$O reaction during the 
periods in-between TPs (interpulse) and by the 
$^{22}$Ne($\alpha,n$)$^{25}$Mg reaction in the 
convective TPs. The $^{13}$C neutron source operates from 
$T \gtrsim 0.9 \times 10^{8}$ K and provides $N_{n} \sim 
10^{7}$ cm$^{-3}$. In order to produce enough $^{13}$C to 
match the enhancements of the $s$-process elements observed in 
AGB stars, it is assumed that some protons are mixed down 
into the He intershell at the deepest inward extent at each TDU 
episode \citep{gallino98,goriely00,busso01}. We refer to the region of the 
He-intershell that has undergone some partial mixing of protons as 
the ``partially mixed zone'' (PMZ). The protons are captured 
by the abundant $^{12}$C producing $^{13}$C and $^{14}$N, 
which allows for the $^{13}$C($\alpha,n$)$^{16}$O reaction to be 
efficiently activated \citep{hollowell90, goriely00,lugaro03a}. 
The $^{22}$Ne neutron source requires 
$T > 3.5 \times 10^{8}$ K to be significantly 
activated and produces $N_{n} \approx 10^{10}$ - 10$^{11}$ 
cm$^{-3}$ in low-mass AGB stars and up to $N_{n} \approx 
10^{13}$ cm$^{-3}$ in intermediate-mass AGB stars 
(see Sect.~\ref{sec:results}). This is because temperatures in excess of 
$3 \times 10^{8}$ K are more easily reached during the TPs 
in intermediate-mass AGB stars \citep{iben75a}. The $^{13}$C 
neutron source is active for a much longer period of time 
($\sim 10^4$ yr) than the $^{22}$Ne neutron source ($\sim 1$ 
yr) and it produces a higher total amount of neutrons, 
even though with a lower neutron density.

\begin{figure}
  \includegraphics[height=.3\textheight]{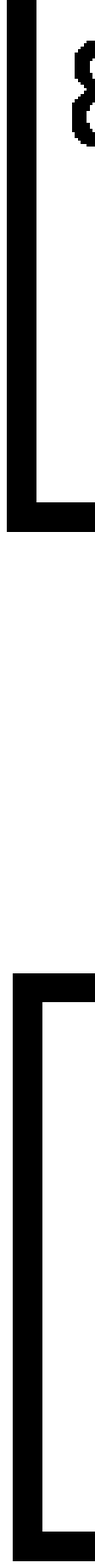}
    \caption{The {\it s}-process path in the region of interest in the chart
    of the nuclides. Isotopes represented by white boxes
    are stable, isotopes represented by black boxes are unstable. 
    The black arrows indicate the {\it s}-process path. The solid arrows
    indicate the {\it s}-process path branches
    specific to low neutron densities. The dashed arrows indicate the
    {\it s}-process path branches followed under high neutron densities. The
    rectangular box encircles isotopes with magic neutron
    number $N=50$. These isotopes have relatively small neutron capture cross sections.}
  \label{fignucbig} \end{figure}

The amount of Rb produced during the $s$ process depends on the 
probability of the two unstable nuclei $^{85}$Kr and $^{86}$Rb 
capturing a neutron before decaying and acting as ``branching points''. 
The probability of this happening depends on the local 
neutron density \citep{beer89}. Figure~\ref{fignucbig} shows a 
section of the chart of the nuclides from Kr to Zr.  
\citet{beer91} reported that, independently of the temperature, 
when $^{84}$Kr captures a neutron, 50\% of the flux 
proceeds to the ground state of $^{85}$Kr (with half-life of 
3934.4 days) and the other 50\% proceeds to the metastable 
state of $^{85}$Kr (with half-life of 4.480 hours). Of the 
metastable $^{85}$Kr, 20\% decays to its ground state while 
the remaining 80\% decays into $^{85}$Rb. Effectively, 
40\% of $^{84}$Kr$+n$ results in the production of 
$^{85}$Rb, whereas the other 60\% results in the production 
of the $^{85}$Kr ground state. The long decay time of the 
$^{85}$Kr ground state allows it to capture another neutron 
to produce $^{86}$Kr instead of $^{85}$Rb, provided the 
neutron density is higher than $\approx 5 \times 10^{8}$ 
cm$^{-3}$. $^{86}$Kr can also capture a neutron, producing 
$^{87}$Kr which quickly decays (with a half-life of 76.3 minutes) 
into $^{87}$Rb. The other unstable isotope in Fig.~\ref{fignucbig} 
is $^{86}$Rb (with a half-life of 18.63 days). This branching 
point produces $^{87}$Rb directly, with $\approx$ 50\% of 
the flux going to $^{87}$Rb if the neutron density is 
$\gtrsim 10^{10}$ cm$^{-3}$ \citep[see also][]{lugaro11}.

In summary, if the branching points at $^{85}$Kr and $^{86}$Rb are 
open $^{87}$Rb is produced. This nucleus has a magic number of 
neutrons of 50 and  a low Maxwellian-averaged neutron 
capture cross-section of 15.7 mbarn at a thermal energy of 
30 keV, as compared to 234 for $^{85}$Rb \citep{heil08}. 
This means that if $^{87}$Rb is produced, it tends to accumulate. 
Therefore, the $^{87}$Rb$/^{85}$Rb isotopic ratios is a direct 
probe of the neutron density at the production site. It is not 
possible to distinguish individual $^{87}$Rb and $^{85}$Rb from 
stellar spectra\footnote{Though they can be measured in the 
interstellar medium \citep{federman04}.} \citep{garcia06,lambert76}.
However, it is possible to determine the elemental Rb abundance 
in relation to other neighbouring {\it s}-process elements, 
for example Sr, Y, or Zr. This ratio is usually expressed relative 
to solar and in a logarithmic scale: [Rb/Zr] or [Rb/Sr]\footnote{$[X/Y] = \mathrm{log}_{10}(X/Y) -
\mathrm{log}_{10}(X_{\sun}/Y_{\sun})$.}. 
A positive/negative [Rb/Zr] or [Rb/Sr] ratio implies a high/low 
neutron density at the {\it s}-process production site. 
This fact has already been used to conclude that the neutron 
density in M, MS, and S stars must be low, and that the 
$^{13}$C($\alpha,n$)$^{16}$O reaction must be the main neutron 
source in these stars \citep{lambert95}.
Theoretical models predict that the activation of the 
$^{22}$Ne neutron source and the maximum value of the neutron 
density are expected to be related to the AGB stellar mass. 
For this reason the [Rb/Zr] ratio in AGB stars is also predicted to be 
positively correlated with the initial mass. Using this 
argument \citet{abia01} concluded that C stars must be AGB 
stars of relatively low mass, $< 3$ \msun.

Intermediate-mass AGB stars over $\gtrsim 4$ \msun\ also 
experience ``hot bottom burning'' (HBB). This occurs during the 
quiescent interpulse phase when the convective envelope extends 
so far inward that it touches the H-burning shell and 
allows proton-capture reactions to take place at the bottom of 
the envelope. The CN cycle is activated, which converts 
$^{12}$C into $^{14}$N and keeping C$<$O in the AGB envelope by 
reducing the C content. Another consequence of HBB is the 
production of lithium (Li) in form of $^{7}$Li via the 
Cameron-Fowler mechanism \citep{cameron77,boothroyd93}. This 
happens via activation of the $^4$He($^3$He,$\gamma$)$^7$Be 
reaction at the base of the envelope, followed by transport 
of $^7$Be towards the stellar surface. There, $^7$Be turns 
into $^7$Li via electron captures while the destruction of 
$^7$Li by proton capture is inhibited by the low temperatures.
Models of intermediate-mass AGB stars find that the
stars become Li-rich during the first $\sim 10$ TPs (Sect.~\ref{sec:li}). 
Hence, an enhanced Li abundance can be interpreted as a signature of the 
AGB star being massive \citep{smith90b}\footnote{Also $\sim$10\% of 
C-rich AGB stars are Li rich \citep{abia93}, which points 
to the occurrence of some kind of ``extra mixing'' below 
the formal border of the convective envelope in low-mass 
AGB stars.}.

Studies of intermediate-mass AGB stars in our Galaxy and 
in the Magellanic Clouds 
\citep[OH/IR stars,][]{garcia06,garcia07,garcia09} have 
yielded the first abundances of Rb, Zr, and Li in these 
stars. The aims of this paper are twofold. First we 
compare our predictions of Rb and Zr production in 
intermediate-mass AGB stars to the observations in order to 
re-confirm that the $^{22}$Ne($\alpha,n$)$^{25}$Mg 
reaction is the main neutron source. Second, we try to 
use the abundances of Rb, Zr, and Li and their observational
trends (e.g., Rb increases with increasing stellar mass and/or 
decreasing metallicity) to constrain the uncertainties in the 
stellar modelling of intermediate-mass AGB stars.
From our detailed comparison we will attempt to define
which stellar modelling uncertainties (or other uncertainties 
e.g., model atmospheres) play the largest role
in the apparent mismatch between observations and theory. 
We will also discuss possible directions for future work 
to study these problems in more detail.

\section{Observations}
\label{sec:obs}

The study of the brightest OH/IR AGB stars in our Galaxy and 
in the Magellanic Clouds by \citet{garcia06,garcia07,garcia09} 
showed that these stars are extremely Rb rich but not Zr rich. 
This suggests that the $^{22}$Ne($\alpha,n$)$^{25}$Mg 
reaction must be the dominant neutron source in intermediate-mass 
AGB stars.

The mass of the observed stars can be estimated from the 
observed Mira pulsation period if the distance to the star is 
known \citep[see, e.g., ][]{wood83}. For the stars in the 
Magellanic Clouds this comparison indicated masses of at least 
6-7 \msun\ \citep{garcia09}. For the observations of Galactic 
stars of \citet{garcia06,garcia07} the distance is unknown. 
However, the Mira pulsation period for these stars shows a 
correlation with the OH expansion velocity ($v_{\mathrm{exp}}$(OH)). 
Independently for this sample, the correspondence of the 
$v_{\mathrm{exp}}$(OH) with mass is obtained using the location of the 
observed stars with respect to the Galactic plane. 
\citep[ See Sect.~5.2 and Sect.~5.3 of][for a full discussion on the 
estimation of mass from $v_{\mathrm{exp}}$(OH)]{garcia07}. From 
the relationship between $v_{\mathrm{exp}}$(OH) and the [Rb/Fe] ratio 
\cite{garcia06} estimated that the [Rb/Fe] ratio increases 
with the initial stellar mass (see their Fig.~2). For the same 
sample of stars \citet{garcia07} also measured the Li abundances 
and found a large variations with $\mathrm{log}($Li$/$H$)+12$ ranging 
from $-$1.0 to 2.6.

In Table~\ref{tab:obs} we summarize the Rb and Zr abundances 
observed in low-mass and intermediate-mass Galactic and 
Magellanic Cloud AGB stars. The putative low-mass stars 
in the Magellanic Cloud reported in Table~\ref{tab:obs}
indicated in brackets are for the stars observed by \citet{plez93} 
in the Small Magellanic Cloud and for the stars classified as 
``HBB-AGB'' in Table 1 of \citet{garcia09}. These stars are 
poor in Rb, but rich in Li, with $\mathrm{log}($Li$/$H$)+12$ 
between $+1.9$ 
and $+3.5$ \citep{plez93}. This composition indicates 
the activation of HBB but not of the $^{22}$Ne neutron source. 
These stars may be more appropriately considered as a
link between low-mass Rb-poor and intermediate-mass Rb-rich 
AGB stars in the Magellanic Clouds. We will discuss these 
stars against model predictions in Sect.~\ref{sec:li}. Table 1 
of \citet{garcia09} also reported two ``Non-HBB-AGB'' 
stars, which have abundances consistent with the values 
reported in Table~\ref{tab:obs} for putative low-mass AGB in 
the Magellanic Clouds, except that one has [Rb/Zr]$>$0.3. 
We note that the distance to the Magellanic Clouds makes even
these brightest of AGB stars faint, thus they are very 
challenging to observe with even the VLT. This means that 
there are only a few observations of intermediate-mass AGB 
stars in the Magellanic Clouds \citep[e.g.,][]{wood83,smith90b,garcia09}. 
Note that in the sample of \citet{garcia09} there are Rb abundances 
for only four stars in the Large Magellanic Cloud and one in the
Small Magellanic Cloud.  
These few stars may represent the first truly massive 
extragalactic AGB stars, as indicated by their strong 
Rb I lines. On the other hand, the Galactic sample of 
intermediate-mass AGB stars is much more 
extensive \citep[120 sources were studied by][]{garcia06,garcia07}. 
The  main trend derived by \citet{garcia09} by comparing 
the Galactic to the Magellanic Cloud stars is that higher 
Rb production is observed in lower metallicity environments, 
with roughly [Rb/Fe]$>$3 in the Large Magellanic Cloud, 
and [Rb/Fe]$<$3 in their Galactic counterparts.

Overall, in low-mass stars the [Rb/Zr] abundance is 
negative, while in intermediate-mass stars it is positive. 
This is in qualitative agreement with the current
theoretical scenario for the $s$ process in AGB stars.
Note that the [Rb/Fe] observed in intermediate-mass AGB stars 
has a large uncertainty of $\pm$ 0.8 dex, which reflects the 
sensitivity of the Rb abundance to changes in the atmospheric 
parameters ($T_{\mathrm{eff}}$, gravity, etc.) adopted for the 
modelling \citep{garcia06}. The observational uncertainties 
allow for an upper limit of [Zr/Fe] $\gtrsim $0.5 for Galactic 
AGB stars, and [Zr/Fe] $\gtrsim $0.3 for the Magellanic Cloud
intermediate-mass AGB stars. By error propagation, 
the [Rb/Zr] ratios have maximum error bars of $\pm$1.0 dex. 
We refer the reader to the literature sources indicated in 
Table~\ref{tab:obs} for more information on the abundance 
derivations and their related error bars.

Serious problems are present in current models of the 
atmospheres of AGB stars. These models are performed in 1D and 
do not include important effects such as the presence of a 
circumstellar dust envelope and dust formation. Furthermore, 
abundance corrections resulting from non local thermodynamic 
equilibrium (NLTE) need to be taken into account. To our 
knowledge, such abundance corrections for Rb do not exist 
in the literature, nevertheless we offer a few comments on 
NLTE. As for most, if not all, NLTE calculations, missing 
atomic data may be an issue for Rb. Although NLTE Rb corrections 
have not yet been established, some expected behaviors may be 
anticipated based on NLTE corrections for other alkali metals 
such as Li, Na, and K \citep[e.g.][]{plez93,asplund05,lind11}. In fact, for 
Li and Na \citet{lind11} noted ``many striking similarities 
between the two elements, especially in the shape of the 
abundance correction curves.'' For Na, the maximum NLTE 
correction is about 0.70 dex. If the NLTE Rb corrections 
mimic the behavior for Li and Na, we may therefore expect (a) 
resonant scattering to be the dominant NLTE mechanism, 
(b) the NLTE correction to be negative, and (c) the magnitude 
of the NLTE correction to reach a maximum for saturated 
lines. The observed Rb abundances in intermediate-mass AGB 
stars come from strongly saturated Rb I lines and
therefore, the NLTE Rb abundances would be expected to be 
smaller, although the exact magnitude of such NLTE Rb 
corrections have yet to be established. We note 
that \citet{plez93} concluded that NLTE Rb effects in the 
stars they analysed are likely to be small.

With regard to the 3D hydrodynamical models, 
the temperature-sensitive Rb I resonance lines originate not 
only from the photosphere of these stars but also in the 
outer nonstatic layers of the stellar atmosphere and in the 
expanding circumstellar shell \citep{garcia06}. Recent years 
have seen a rapid development of numerical 3D radiative
hydrodynamic simulations of stellar surface convection 
\citep[see e.g.,][and references therein]{collet11}.
In these simulations,  gas flows in the highly stratified 
outer layers of stars are modelled by solving the hydrodynamic
equations of mass, momentum, and energy conservation and 
accounting for the energy exchanges between matter and 
radiation via radiative transfer. In this framework, 
convective motions arise and self-organize naturally without 
the need to introduce adjustable parameters. One of the main 
goals of 3D modelling of stellar surface convection is 
therefore to provide a more realistic description of the 
physical structure of late-type stellar atmospheres. From the 
point of view of spectroscopic analyses, the structural
differences between 3D and hydrostatic (1D) models can lead 
to large differences in terms of derived abundances whenever 
temperature-sensitive spectral features are used \citep{asplund05}. 
In short, 3D modelling shows that 3D effects are more important at 
low metallicity and low gravities. This means that the 
Rb abundances obtained from the sensitive-temperature Rb I 
resonance lines would be higher in the hydrostatic models 
compared with the 3D hydrodynamical ones, as a consequence 
of the higher temperature in the static case, since the 
strength of the Rb I resonance lines increases with 
decreasing temperature \citep{garcia06}. We would expect 
drastic 3D effects in intermediate-mass AGB stars at low metallicity. 
In summary, possible NLTE and 3D Rb corrections could decrease the 
Rb abundances derived from hydrostatic LTE models by (at least) 
one order of magnitude. However, these speculations need to be 
confirmed by detailed 3D hydrodynamical simulations and NLTE calculations.

\begin{center} 
\begin{table*} 
\caption{Ranges of the observed Rb and Zr abundances in AGB stars in the Galaxy and in the 
Magellanic Clouds.} 
\label{tab:obs} 
\begin{tabular}{ c c c c c } \hline
  & \multicolumn{2}{ c }{Putative low-mass AGB stars} &  
\multicolumn{2}{ c }{Putative intermediate-mass AGB stars} \\
  & Galaxy & Magellanic Clouds\tablefootmark{a} & Galaxy & Magellanic Clouds \\ 
\hline

[Rb/Fe\tablefootmark{b}]     & $-0.3$ - $+0.9$ & ($-1.2$ - $+0.4$)   & $+0.4$ - $+2.5$ & $+1.7$ - $+5.0$ \\

$[$Zr\tablefootmark{c}/Fe\tablefootmark{b}] & $-0.3$ - $+1.4$ & ($-0.3$ - $+0.70$)  & 0\tablefootmark{d} & 0\tablefootmark{d} \\

[Rb/Zr] & $<$ 0 & ($<$ 0) & $>$ 0 & $>$ 0 \\

[Fe\tablefootmark{b}/H] & $-1.3$ - $+0.1$ & ($-0.70$ - $-0.37$\tablefootmark{e}) & set to 0\tablefootmark{e}
& 
set to $-0.3$\tablefootmark{e} and $-0.7$\tablefootmark{e} \\ 

\hline
  & \citet{lambert95} & \citet{plez93} & \citet{garcia06} & \citet{garcia09} \\
  & \citet{abia01} & \citet{garcia09} & \citet{garcia07} & \\
\hline 
\end{tabular}
\tablefoot{
\tablefoottext{a}{As discussed in the text, these stars have high Li abundances and
may represent a link between low-mass and intermediate-mass AGB stars in the Magellanic Clouds.}
\tablefoottext{b}{In some of the reported studies the average of Fe and Ni, M, is used instead of Fe.}
\tablefoottext{c}{In \citet{lambert95} [$s$/M] is reported as the average of
several $s$-process elements. We use this value as an 
indicator of the [Zr/Fe] ratio for stars from this study.}
\tablefoottext{d}{The observational uncertainties allow for an upper limit of
$+0.5$ and of $+0.3$ for Galactic and Magellanic Clouds 
intermediate-mass AGB stars, respectively.}
\tablefoottext{e}{The metallicity for each star is not reported and the  
Rb and Zr abundances are derived using model atmospheres for the average abundances [M/H] 
set to 0, $-$0.3, and $-$0.7 in the Galaxy, Large and Small Magellanic Clouds, respectively.}
}
\end{table*} 
\end{center}

\section{Stellar models} \label{sec:models}

We calculate the nucleosynthesis with a detailed post-processing code
\citep{cannon93} for which the stellar evolution 
inputs were calculated beforehand. The stellar evolution models are 
from \citet{karakas07b} and were calculated with the 
Monash version of the Mount Stromlo Stellar Structure Program 
\citep[see][and references therein for details] {frost96} 
with AGB mass loss based on the formula by \citet{vw93}. 
The choice of mass loss is one of the most important 
uncertainties in the computation of stellar evolution and it affects 
the nucleosynthesis by determining the stellar lifetime, which in 
turn changes the amount of TDU episodes as well as the timescale for 
the operation of HBB. Another important stellar model uncertainty is 
the treatment of convection and of convective borders. In our code we use the 
mixing length theory to parametrize convection with the free parameter
$\alpha$ set to the value of 1.75. A different parametrization of
convection and a different choice of the mass loss have been shown to
have an important impact on the operation of HBB and the resulting 
AGB yields \citep{ventura05a,ventura05b}. Furthermore, the treatment of convective 
boundaries affects the efficiency of the TDU \citep[e.g.][]{frost96}. 
Here, we apply the Schwatzschild criterion and then 
extend the convective zone to the neutral border in the same way as described in 
\citet{lattanzio86}.

The post-processing code calculates the abundances due to convective
mixing and nuclear reaction rates for a large number 
of species. We use a nuclear network of 166 species from neutrons and 
protons to S and from Fe to Nb. The missing species 
are accounted for by means of artificial neutron sinks. For one 
specific model, the 6 \msun\ $Z=0.02$, we tested the 
results produced using a full nuclear network of 320 species and no 
neutron sinks and found no differences (to within 0.1 dex) for
[Rb/Fe], [Zr/Fe], and [Rb/Zr]. The bulk of our 1285 reaction rates 
are from reaclib \citep{thielemann86}, with updated 
tables as described by \citet{lugaro04} and \citet{karakas06a} in 
particular for the $^{22}$Ne($\alpha,n$)$^{25}$Mg 
reaction. We further updated the neutron capture rates to those of \citet{bao00}.

We have used here the same stellar models presented by
\citet{lugaro07} to compare to the composition of peculiar 
stardust spinel grain OC2 and by \citet{karakas09} to compare to 
the observations of heavy elements in planetary nebulae 
of Type I. The $Z=0.008$ and $Z=0.004$ models were also already 
used by \citet{garcia09} to compare to the observations 
of intermediate-mass Rb-rich O-rich AGB stars in the Magellanic Clouds.

A summary of the properties of the models is given in
Table~\ref{tab:models}, including the mass (Mass) and metallicity 
($Z$) of the model, the number of TPs (TPs), the maximum temperature 
achieved in the TPs ($T^{\mathrm{max}}_{\mathrm{He}}$), the maximum temperature 
achieved at the base of the convective envelope ($T^{\mathrm{max}}_{\mathrm{bce}}$), 
the total mass dredged-up to the surface by the TDU ($M_{\mathrm{dred}}$), 
the final envelope mass at the end of the computed evolution 
($M_{\mathrm{env}}$), if HBB is at work, and the final C/O and 
$^{12}$C/$^{13}$C ratios.

The observations of \citet{garcia06} and \citet{garcia07} are of 
Galactic-disk AGB stars, and for this reason we 
concentrated on models of solar metallicity $Z=0.02$ for masses 
3, 4, 5, 6, and 6.5 \msun. The initial abundances 
were taken from \citet{anders89}. We also calculated a 6.5 \msun\
model using the solar abundances from \citet{asplund05} 
and \citet{lodders03}, which prescribe a solar metallicity of 
$Z=0.012$. We further calculated low metallicity 
models of 6 \msun \space $Z=0.008$, 5 \msun \space 
with $Z=0.008$, and 4 and 5 \msun \space with $Z=0.004$ which are suitable for 
comparison to the Magellanic Cloud AGB stars. For the 
low-metallicity models we used the \citet{anders89} 
abundances scaled down to the appropriate $Z$.

In four selected models we artificially included a PMZ in the top 
layers of the intershell in the same way described in 
detail by \citet{lugaro04}. For the 3 \msun\space model we used a 
PMZ of size $2 \times 10^{-3}$ \msun\space. For the 4, 
5, and 6.5 models we used a PMZ of size $1 \times 10^{-4}$
\msun\space. This smaller value is because the mass of the 
intershell is about one order of magnitude smaller in the higher 
than in the lower mass AGB stars ($\sim 10^{-3}$ \msun\ 
with respect to $\sim 10^{-2}$ \msun), so we scaled the mass of 
the PMZ accordingly. The inclusion of a PMZ is the 
main uncertainty in models of the $s$ process. We still do not 
know the mechanism and the features of the process driving 
the mixing, although some promising candidates have been 
proposed \citep[see][for a discussion]{busso99}. 
Stellar models of intermediate-mass AGB stars indicate that any 
protons mixed into the He-intershell during the TDU would likely
burn to $^{14}$N before producing $^{13}$C owing to the high
temperatures at the base of the envelope
\citep{siess04,herwig04a}. This would inhibit the release of neutrons
by the $^{13}$C neutron source. For this reason, the artificial 
inclusion of a PMZ into a model of an intermediate-mass AGB star
should be treated with some caution.

\begin{center} 
\begin{table*}[ht] 
\begin{threeparttable} \caption{Details of stellar models.}
  \label{tab:models}
  \begin{tabular}{ c c c c c c c c c c }
    \hline
    Mass & $Z$ & TPs\tablefootmark{a} & $T^{\mathrm{max}}_{\mathrm{He}}$ & 
$T^{\mathrm{max}}_{\mathrm{bce}}$ & $M_{\mathrm{dred}}$ & 
final $M_{\mathrm{env}}$ & HBB & C/O & $^{12}$C/$^{13}$C \\
     & & & (MK) & (MK) & (\msun) & (\msun) & & & \\
    \hline
    \hline
    3 \msun & 0.02 & $26+0$ & 302 & 6.75 & 0.081 & 0.676 & No & 1.40 & 117.47 \\
    4 \msun & 0.02 & $18+2$ & 332 & 22.7 & 0.056 & 0.958 & No & 0.99 & 76.22 \\
    5 \msun & 0.02 & $24+5$ & 352 & 64.5 & 0.050 & 1.500 & Yes & 0.77 & 7.83 \\
    6 \msun & 0.02 & $438+5$ & 369 & 83.1 & 0.058 & 1.791 & Yes & 0.38 & 10.65 \\
    6.5 \msun & 0.02 & $40+7$ & 368 & 86.5 & 0.047 & 1.507 & Yes & 0.34 & 9.76 \\
    6.5 \msun & 0.012 & $52+2$ & 369 & 90.0 & 0.065 & 1.389 & Yes & 0.75 & 10.40 \\
    5 \msun & 0.008 & $59+4$ & 366 & 80.8 & 0.17 & 1.795 & Yes & 0.93 & 7.49 \\
    6 \msun & 0.008 & $69+5$ & 374 & 89.6 & 0.12 & 1.197 & Yes & 1.41 & 8.88 \\
    4 \msun & 0.004 & $30+2$ & 366 & 66.3 & 0.11 & 1.333 & Yes & 3.18 & 7.98 \\
    5 \msun & 0.004 & $83+3$ & 377 & 84.4 & 0.22 & 1.141 & Yes & 2.59 & 7.88 \\
    \hline
    \hline
  \end{tabular}
\tablefoot{
\tablefoottext{a}{$+n$ refers to the number of synthetic TPs as described in
     Sect.~\ref{sec:extended}}
}
  \end{threeparttable}
  \end{table*} \end{center}

\subsection{Late AGB-phase evolution} \label{sec:extended}

Owing to convergence difficulties, the stellar evolution models used
as input into the post-processing calculations were 
not evolved through to the end of the superwind phase\footnote{The 
superwind is the final stage of the AGB phase when very high mass
loss, up to $10^{-4}$ \msun/yr \citep{iben+renzini83} may occur.} 
and into the post-AGB phase. That the stellar models have some
remaining envelopes (see Table~\ref{tab:models}) means that there 
could be further TPs and TDU events \citep{karakas07b}. These 
latest stages of AGB star evolution are pivotal to the abundances
under study here. Not only are the observed AGB stars in their late 
stages of evolution, but the freshly synthesized isotopes from the stellar 
interior are less diluted inside the rapidly decreasing envelope. 
To calculate the surface abundance evolution during the 
final stages of AGB evolution, we use a synthetic patch that we run
after the nucleosynthesis code has finished. We 
estimate the remaining number of pulses based on the stellar
parameters during the final time step that was calculated by 
the stellar evolution code, as described in detail by
\citet{karakas07b}.  The number $n$ of remaining pulses for each 
model are reported in Table \ref{tab:obs} as $+n$ in Column 3. 
For example, we estimate that 5 and 7 more TP will occur 
for the 6 \msun\space and 6.5 \msun, $Z=0.02$ models, respectively.

The most crucial component for our purposes in these calculations is
the TDU efficiency as a function of the decreasing 
envelope mass. The TDU efficiency $\lambda$ is defined as the fraction
of the mass that is dredged up over the mass by 
which the H-exhausted core had increased in the previous interpulse 
period, i.e., $\lambda = \frac{\Delta M_{\mathrm{DUP}}}{\Delta M_{\mathrm{core}}}$. The 
TDU efficiency in these final phases when the envelope is decreasing 
could range from $\lambda = 0$ \citep{straniero97} 
to $\lambda_{\mathrm{max}} \approx 0.9$, at the last TP
\citep{stancliffe07b,karakas10a}. In our case these (maximum)
efficiencies are: $\lambda = 0.93$, 0.95,
and 0.94, for models of 5, 6, and 6.5 \msun, respectively. This 
simple synthetic extension of the evolution does not take into account
the effect of HBB. However in all cases HBB has ceased in the detailed
stellar evolution calculations. The Li abundance is not expected to vary 
significantly at this stage, owing to the depletion of the $^3$He fuel
needed for Li production and the cessation of HBB.

\begin{figure}
\includegraphics[height=.35\textheight,angle=-90]{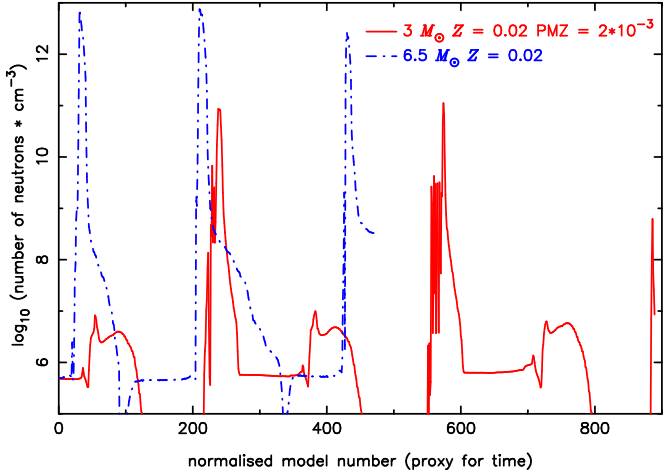}
\caption{Maximum neutron density in the He intershell versus normalised model
number (proxy for time) showing the last three TPs computed for a 3
\msun\space AGB star with a PMZ of 2 $\times 10^{-3}$ \msun\space
(red curve) and a 6.5 \msun\space model without a PMZ included (blue
curve). The peaks are a result of the $^{22}$Ne($\alpha,n$)$^{25}$Mg
reaction activated during TPs. For the 3 \msun\space model the neutron flux during each TPs lasts 
for $\sim$ 4 yr and the total time-integrated neutron flux (neutron exposure $\tau$) is $\sim 0.02$ 
mbarn$^{-1}$. In the 6.5 \msun\space model the neutron density reaches $10^9$ cm$^{-3}$ for 
$\sim 3$ yr and $\tau \sim 0.2$ mbarn$^{-1}$. The secondary bumps
at $\sim 10^{7}$ cm$^{-3}$ in the 3 \msun\space model are a result of the $^{13}$C($\alpha,n$)$^{16}$O
reaction, which is activated during the interpulse phase due to the PMZ. This activation lasts for 
$\sim 30,000$ yr and results in $\tau \sim 0.4$ mbarn$^{-1}$ 
\citep[see also][]{lugaro03a}. The plateaux at $\sim 10^{6}$ cm$^{-3}$ represent neutrons 
released during radiative shell He burning in the interpulse periods. 
These neutron fluxes are insignificant  
compared to the others (with $\tau <$ 0.02 mbarn$^{-1}$). Furthermore, the
material affected does not make it to the stellar surface as it remains buried in 
the C-O core. (Colour figure available online).}
\label{neutrondensity} 
\end{figure}

\begin{center} \begin{table*}[ht] \rowcolors{1}{white}{tableShade}
  \caption{Final envelope abundances and ratios involving Rb, Zr, the isotopes of Rb, and Li 
for all our computed models.} 
  \begin{tabular}{ c c c | c c c c c c }
      \hline
          $M$ & $Z$ & PMZ mass (\msun) & Pulse & $[$Rb$/$Fe$]$ & $[$Zr$/$Fe$]$
	  & $[$Rb$/$Zr$]$ & $[^{87}$Rb$/^{85}$Rb$]$ & $\mathrm{log}($Li$/$H$)+12$ \\ \hline
	  \hline
	  3 \msun & 0.02 & 0 & 26 & 0.00 & 0.00 & $-0.01$ & 0.00 & 1.51 \\
	  3 \msun & 0.02 & $2 \times 10^{-3}$ & 26 & 0.50 & 1.00 & $-0.49$ & $-0.11$ & 1.51 \\
	  4 \msun & 0.02 & 0 & 18 & 0.01 & 0.01 & 0.00 & 0.01 & 1.30 \\
	  \multicolumn{3}{ c | }{$extended$} & 20 & 0.03 & 0.03 & 0.00 & 0.02 & \\
	  4 \msun & 0.02 & $1 \times 10^{-4}$ & 18 & 0.57 & 0.79 & $-0.22$ & 0.08 & 1.18 \\
	  \multicolumn{3}{ c | }{$extended$} & 20 & 0.83 & 1.21 & $-0.38$ & 0.15 & \\
	  5 \msun & 0.02 & 0 & 24 & 0.05 & 0.01 & 0.03 & 0.07 & 2.27 \\
	  \multicolumn{3}{ c | }{$extended$} & 29 & 0.33 & 0.16 & 0.17 & 0.37 &  \\
	  5 \msun & 0.02 & $1 \times 10^{-4}$ & 24 & 0.39 & 0.40 & $-0.01$ & 0.37 & 2.37 \\
	  \multicolumn{3}{ c | }{$extended$} & 29 & 0.80 & 0.84 & $-0.04$ & 0.63 &  \\
	  6 \msun & 0.02 & 0 & 38 & 0.21 & 0.07 & 0.14 & 0.25 & 1.06 \\
	  \multicolumn{3}{ c | }{$extended$} & 43 & 0.73 & 0.46 & 0.27 & 0.61 &  \\
	  6.5 \msun & 0.02 & 0 & 40 & 0.26 & 0.09 & 0.16 & 0.30 & 0.93 \\
	  \multicolumn{3}{ c | }{$extended$} & 47 & 1.04 & 0.73 & 0.31 & 0.69 &  \\
	  6.5 \msun & 0.012 & 0 & 52 & 0.22 & 0.07 & 0.15 & 0.25 & $-0.49$ \\
	  \multicolumn{3}{ c | }{$extended$} & 54 & 0.69 & 0.36 & 0.33 & 0.56 &  \\
	  6.5 \msun & 0.012 & $1 \times 10^{-4}$ & 52 & 0.62 & 0.60 & 0.02 & 0.59 & $-1.03$ \\
	  \multicolumn{3}{ c | }{$extended$} & 54 & 1.04 & 1.04 & 0.00 & 0.76 &  \\
	  5 \msun & 0.008 & 0 & 59 & 0.71 & 0.34 & 0.37 & 0.52 & $-0.33$ \\
	  \multicolumn{3}{ c | }{$extended$} & 63 & 1.30 & 1.02 & 0.29 & 0.76 &  \\
	  6 \msun & 0.008 & 0 & 69 & 0.90 & 0.51 & 0.39 & 0.60 & $-0.50$ \\
	  \multicolumn{3}{ c | }{$extended$} & 74 & 1.44 & 1.10 & 0.34 & 0.77 &  \\
	  4 \msun & 0.004 & 0 & 30 & 0.80 & 0.28 & 0.52 & 0.46 & 2.98 \\
	  5 \msun & 0.004 & 0 & 83 & 1.12 & 1.08 & 0.04 & 0.21 & $-2.68$ \\
	  \multicolumn{3}{ c | }{$extended$} & 86 & 1.38 & 1.45 & $-0.08$ & 0.26 & \\
        \hline
	\hline
    \end{tabular}
     \tablefoot{Rows are alternately coloured for ease of
reading. The rows labelled as {\it extended} report the final results obtained using 
the simple synthetic extension of the evolution described in Sect.~\ref{sec:extended}}.
    \label{tabresults} \end{table*} \end{center}

\begin{center} \begin{table*}[ht] \rowcolors{1}{white}{tableShade}
  \caption{Same as Table~\ref{tabresults} but for abundances and
ratios involving C, N, O, Na, Mg and the isotopes of Mg 
for our computed models without inclusion of a PMZ.}.
  \begin{tabular}{ c c | c c c c c c c }
      \hline
          $M$ & $Z$ & $[$C$/$Fe$]$ & $[$N$/$Fe$]$
	  & $[$O$/$Fe$]$ & $[$Na$/$Fe$]$ & $[$Mg$/$Fe$]$ & $[^{25}$Mg$/^{24}$Mg$]$ 
	  & $[^{26}$Mg$/^{24}$Mg$]$ \\ 
         \hline
	  \hline
	  3 \msun & 0.02 & 0.49 & 0.34 & $-$0.02 & 0.19 & 0.01 & 0.03 & 0.03 \\
	  4 \msun & 0.02 & 0.34 & 0.38 & $-$0.02 & 0.17 & 0.03 & 0.10 & 0.10 \\
	  5 \msun & 0.02 & 0.21 & 0.57 & $-$0.04 & 0.28 & 0.04 & 0.13 & 0.18 \\
	  \multicolumn{2}{ c | }{$extended$} & 0.60 & 0.56 & $-$0.05 & 0.31 & 0.14 & 0.41 & 0.50 \\
	  6 \msun & 0.02 & $-$0.14 & 0.89 & $-$0.08 & 0.85 & 0.07 & 0.22 & 0.31 \\
	  \multicolumn{2}{ c | }{$extended$} & 0.46 & 0.88 & $-$0.08 & 0.85 & 0.20 & 0.51 & 0.67 \\
	  6.5 \msun & 0.02 & $-$0.19 & 0.87 & $-$0.09 & 0.79 & 0.06 & 0.22 & 0.30 \\
	  \multicolumn{2}{ c | }{$extended$} & 0.70 & 0.86 & $-$0.09 & 0.80 & 0.32 & 0.69 & 0.89 \\
	  6.5 \msun & 0.012 & $-$0.16 & 0.87 & $-$0.41 & 0.64 & 0.04 & 0.46 & 0.38 \\
	  5 \msun & 0.008 & 0.19 & 1.51 & $-$0.14 & 1.52 & 0.31 & 0.59 & 0.77 \\
	  6 \msun & 0.008 & 0.26 & 1.39 & $-$0.25 & 1.04 & 0.28 & 0.79 & 0.80 \\
	  4 \msun & 0.004 & 0.88 & 1.57 & $-$0.17 & 0.93 & 0.35 & 0.68 & 0.92 \\
	  5 \msun & 0.004 & 0.56 & 1.89 & $-$0.22 & 1.78 & 0.63 & 0.83 & 1.10 \\
        \hline
	\hline
    \end{tabular}
\tablefoot{Solar reference ratios are taken from \citet{anders89}.}
    \label{tabresultslight} \end{table*} \end{center}

\section{Results} \label{sec:results}

Figure \ref{neutrondensity} shows the maximum neutron density in the He intershell calculated for two of 
our models, a low-mass and a intermediate-mass AGB star, as a function of normalised model number (proxy 
for time). As anticipated, the maximum neutron density from the activation of the $^{22}$Ne neutron 
source is up to two orders of magnitude higher in the intermediate-mass AGB than in the low-mass AGB 
model, where instead the $^{13}$C is the main neutron source. This results in positive and negative 
[Rb/Zr] in the two models, respectively.

\begin{figure}

\includegraphics[height=.35\textheight,angle=-90]{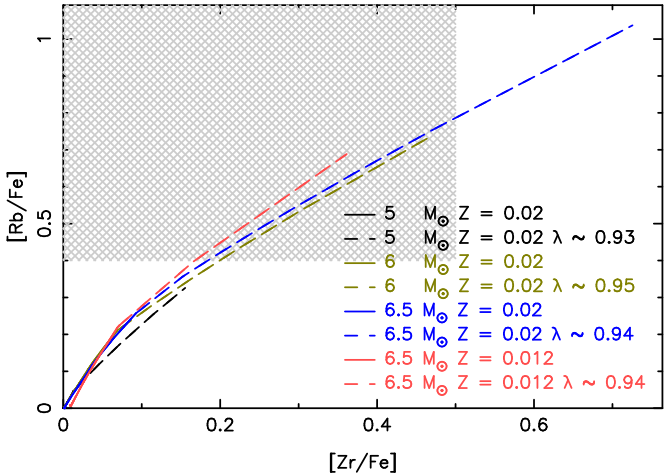}
    \caption{Model results for the [Rb/Fe] against [Zr/Fe] for
    intermediate-mass AGB stars of solar metallicity
      without the inclusion of a PMZ. The solid part of the lines
      are equivalent to a TDU efficiency of 0 during the extended evolution.
      The dashed lines are the result of the synthetic extension when the
      TDU efficiency is chosen to remain constant. The observed [Rb/Fe]
    and [Zr/Fe] ranges are indicated by the grey shaded area, note
    that [Rb/Fe] ranges up to 2.5 ($\pm$ 0.8, see 
    Table~\ref{tab:obs}). (Colour figure available online).}
      \label{rbfehm} \end{figure}

\begin{figure}
  \includegraphics[height=.35\textheight,angle=-90]{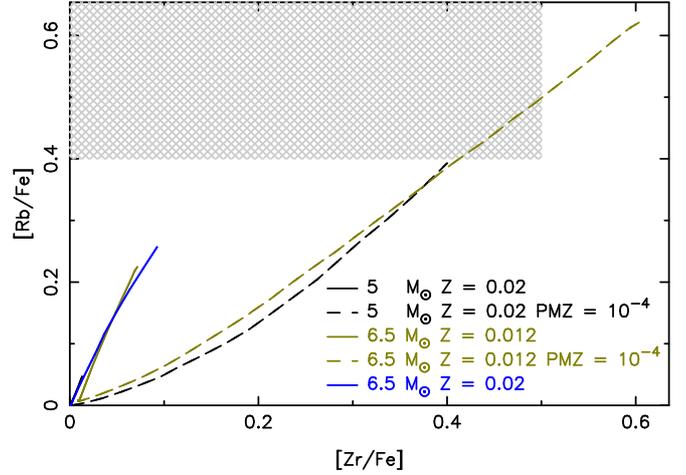}
      \caption{Results for selected intermediate-mass AGB models with
     and without the inclusion of PMZs. The solid lines 
      represent
      model predictions without the inclusion of a PMZ, the
      dashed line models with a PMZ of $1 \times 10^{-4}$ \msun. The observed [Rb/Fe]
    and [Zr/Fe] ranges are indicated by the grey shaded area, 
    note that [Rb/Fe] ranges up to 2.5 ($\pm$ 0.8, see 
     Table~\ref{tab:obs}). (Colour figure available online).}
      \label{rbzrhmpmz} \end{figure}

Our results for Rb, Zr, and Li are summarised in
Table~\ref{tabresults} and illustrated in Figs.~\ref{rbfehm}, 
\ref{rbzrhmpmz}, \ref{figdave}, and \ref{livspulsenumber}. As reference
for future observations, in Table~\ref{tabresultslight} we present the 
results for a set of key light elements (C, N, O, Na, and Mg) and the
Mg isotopic ratios. Inclusion of a PMZ does not significantly affect
the results for the light elements, except for the [Na/Fe] ratio in the 3 \msun\ and 4 
\msun\ models with $Z=0.02$, which increases by roughly $+0.14$ dex with the inclusion of the PMZ 
\citep[see also][]{goriely00}. Same examples of the effect on the
light elements of the synthetic extension of the models are also
reported in Table~\ref{tabresultslight}.  The yields of elements
up S for a large range of stellar models, including those presented
here, have been published in \citet{karakas10a} and results for 
Zn, Ge, Se, Br, and Kr are published in \citet{karakas09}. 

We predict an increase in [Rb/Fe] with increasing stellar mass, as is
implied by the observations. This is because a higher stellar
mass leads to (i) hotter temperatures during He-shell burning and 
a stronger activation of the $^{22}$Ne neutron source. This results in $s$-process
enhancements of [Rb,Zr/Fe] in models more massive than 4 \msun\ 
even without the inclusion of the PMZ. Also, (ii) a higher number of TPs
with high He-burning temperatures and followed 
by efficient TDU episodes (Table~\ref{tab:obs}). However, 
intermediate-mass AGB models without synthetic extensions or 
PMZs produce [Rb/Fe] ratios that do not match the observed values, 
while the predicted [Zr/Fe] ratios are within the 
observed range ($< 0.5$, Fig.~\ref{rbfehm}).

We included a PMZ of $1 \times 10^{-4}$ \msun\space in our
calculations for the 5 \msun, $Z=0.02$ and 6.5 \msun, $Z=0.012$ 
models and found that [Rb/Fe] increases by a factor of 2 to 3. 
However, [Zr/Fe] also increases by similar factors. This 
results in [Rb/Zr]$\sim 0$, lower than found without the 
inclusion of a PMZ (Fig. \ref{rbzrhmpmz}). A PMZ in 
intermediate-mass AGB stars may not provide a solution to the 
Zr and Rb abundances investigated here.

In comparison with the 5 \msun\ models presented by \citet{abia01}, 
we obtain lower production factors for Rb and Zr. 
This may be due to the different choice of mass-loss rate. 
\citet{abia01} used the prescription by \citet{reimers75} with 
a choice of the free parameter $\eta=10$, while we used the 
prescription of \citet{vw93}. However, Table 1 of \citet{straniero00} 
indicates that the 5 \msun\ models used in \citet{abia01} experienced
23 TPs, very close to the 
number in our model. The difference must be due to a different
$^{13}$C abundance in their PMZ. 
From comparison of the low-mass 
models we derive that the results from our 3 \msun\space AGB model 
are in agreement with the 1.5 \msun\ model presented 
by \citet{abia01} when their standard choice of the amount of the 
$^{13}$C neutron source is selected. For their 5 \msun\ 
model, \citet{abia01} divide the amount of $^{13}$C by a factor of 
10, while we divide the extent in mass of the PMZ by a 
factor of 20, which results in lower abundances. In any case, there 
is agreement in terms of the [Rb/Zr] ratio both 
specifically for the 5 \msun\ model, as well as in general, since 
our intermediate-mass AGB models all produce positive 
values ratios up to $+0.4$ (except in the 4 \msun\, $Z=0.02$ model with a 
PMZ included and the 5 \msun, $Z=0.004$ model), 
similarly to the models of \citet{abia01} (except for 
one negative value of $-$0.4 obtained for their specific 
standard$\times$2 choice of the $^{13}$C amount).

\subsection{Synthetic Extension of the Models}

\label{secsynthextendresults}

We extended our models to evolve through the superwind phase using the
synthetic algorithm described in Sect.~\ref{sec:extended}. The resulting 
abundances are plotted in Fig.~\ref{rbfehm} (dashed parts of the
lines). The solid parts of the curves in Fig.~\ref{rbfehm} are
equivalent to assuming $\lambda=0$ during the synthetic extension of the 
models, while the dashed parts of the curves show the results of
extending the models using the same dredge up efficiency 
as during the last computed pulse, i.e., the maximum possible dredge
up efficiency. In this case, we find sufficiently 
high [Rb/Fe] at the final pulse for the 6 \msun\ and 6.5 \msun\ 
models to match the lowest range of the observations 
(within the observational error bar). We re-emphasise that this 
synthetic extension is a rough estimate. Even in this 
case we still cannot match the high end of the observed [Rb/Fe] 
values, i.e., $>$ 1.8. These may be accounted for by AGB 
stars even more massive than 6.5 \msun. Another difficulty is that 
most likely HBB is shut off in these late phases and 
hence it is not possible to keep the star oxygen rich as observed 
because carbon is carried to the surface via the TDU 
but it is not consumed by proton captures at the base of the 
envelope (Table~\ref{tabresultslight}).

\subsection{Reaction Rates}

Using the 6.5 \msun\ and 6 \msun\ $Z=0.02$ models we performed several
tests changing the neutron-capture cross sections 
involved in the production of Rb specifically with the aims of (i) 
investigating the relative impact of the two branching 
points at $^{85}$Kr and $^{86}$Rb on the production of Rb and (ii) 
finding the choice of neutron-capture rates, within 
their uncertainties, that would maximise the Rb production. The 
neutron-capture cross sections of the unstable isotopes 
$^{85}$Kr and $^{86}$Rb are evaluated only theoretically, and 
the estimates vary roughly by a factor of two (see the 
$www.kadonis.org$ database) thus we multiplied and divided them 
by a factor of 2. The neutron-capture cross section of 
$^{86}$Kr has been measured both via the activation technique 
and the Time Of Flight (TOF) technique. The rate 
recommended by \citet{bao00} is based on the activation 
measurements, however, the TOF measurements are typically higher 
(up to 70\%) than the activation measurements. We made a test 
multiplying the $^{86}$Kr($n,\gamma)^{87}$Kr cross 
section by a factor of 2. The neutron-capture cross sections 
of $^{85}$Rb and $^{87}$Rb have been recently re-measured by 
\citet{heil08}, who confirmed the values of the previous experiments 
with a high precision.  We also made tests with some 
combinations of the above choices. The overall result is that 
the [Rb/Zr] ratio increased at most by 0.28 dex, when 
considering the synthetic extensions of the models and the 
following combination of test neutron-capture cross sections 
$\sigma$: $\sigma$[$^{85}$Kr$(n,\gamma)^{86}$Kr]/2, 
$\sigma$[$^{86}$Kr$(n,\gamma)^{87}$Kr]$\times$2, and 
$\sigma$[$^{86}$Rb$(n,\gamma)^{87}$Rb]$\times$2.

Given the surprising result that more Rb was produced when the 
efficiency of the $^{85}$Kr branching factor was smaller, 
we decided to test the differential impact of the $^{85}$Kr and 
the $^{86}$Rb branching points on the production of Rb 
using the 6 \msun\ $Z=0.02$ model. Assuming instantaneous 
$\beta$-decay of $^{85}$Kr we obtained [Rb/Fe] $+0.14$ 
higher because [$^{87}$Rb/Fe] decreased by $-$0.08 dex, but 
[$^{85}$Rb/Fe] increased by $+0.25$ dex. We ascribe this 
result to the fact that when the $^{85}$Kr branching point is 
open $^{86}$Kr is produced, which has an extremely low 
Maxwellian-averaged neutron-capture cross section of 3.4 mbarn 
at a thermal energy of 30keV \citep{beer91}, and 
it accumulates at the expenses of $^{85}$Rb and $^{87}$Rb. 
In fact, [$^{86}$Kr/Fe] decreased by $-$0.24 dex in this 
test case. Assuming instantaneous $\beta$-decay of 
$^{86}$Rb we obtained instead [Rb/Fe] $-$0.08 lower as a 
consequence of a $-$0.25 dex lower [$^{87}$Rb/Fe].

Finally, again using the 6 \msun\ $Z=0.02$ model, we checked the effect of varying the 
$^{22}$Ne($\alpha,n$)$^{25}$Mg reaction rate. When varying the rate
between the upper and lower limits described in 
\citet{karakas06a} we found variations in [Rb/Zr] of $+0.16$ dex at
most. When using the rate estimated in the 
NACRE compilation \citep{angulo99}, which is up 40\% faster than that 
of \citet{karakas06a} at typical He-burning temperatures, we found an 
increase of $+0.21$ dex in [Rb/Fe] and of $+0.07$ dex increase in [Zr/Fe], resulting in a 
$+0.14$ dex increase in the [Rb/Zr]. We have also computed 
selected models using the new $^{22}$Ne$ + \alpha$
reaction rates from \citet{iliadis10}. The main finding is that with
the new rates both [Rb/Fe] and [Zr/Fe] can increase by up to roughly $+0.2$ dex,
in the 6 \msun\ $Z=0.008$ model, which is one of the models with the highest
He-burning temperatures. For the other models the
effect is smaller and basically negligible at solar metallicity. The
reasons are that (i) the new $^{22}$Ne($\alpha,n$)$^{25}$Mg rate is 
not very different from that used in our study and (ii) the fact that 
the revised $^{22}$Ne($\alpha,\gamma$)$^{26}$Mg rate is roughly an order
of magnitude lower than that used here is not significant for the $s$ 
process. This is because the amount of $^{22}$Ne
that burns is very small in any case: at most the $^{22}$Ne intershell
abundance decreases by a factor of two. Hence, that the two
$\alpha$-capture channels do not need to compete with each
other for the $^{22}$Ne.

\subsection{Low-metallicity models} \label{seclowmetal}

\begin{figure} \begin{center} 
\includegraphics[height=.35\textheight,angle=-90]{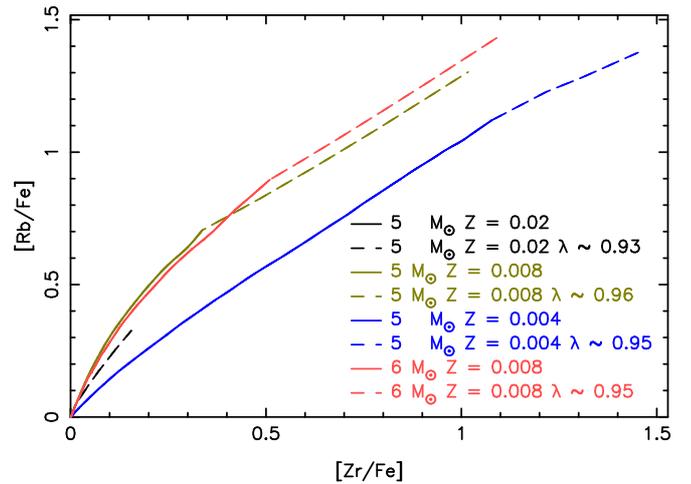}
  \caption{ Resulting [Rb/Fe] and [Zr/Fe] for low-metallicity intermediate-mass AGB
  models indicated by the labels. For comparison we included the 5
  \msun \space $Z=0.02$ model. As in Fig.~\ref{rbfehm} the solid part of the lines
	  are equivalent to $\lambda=0$ during the extended evolution.
	  The dashed lines are the result of the synthetic extension
	  when the TDU efficiency is chosen to remain constant. The
          observed [Rb/Fe] in five Rb-rich stars in the 
	  Magellanic Clouds ranges from $+1.7$ dex to $+5.0$ dex, 
    	  outside of the range of the figure. (Colour figure available online).} 
\label{figdave} \end{center} 
\end{figure}

Fig.~\ref{figdave} shows that low-metallicity intermediate-mass 
AGB stars show significant enhancements of Rb and Zr, 
even without including the final synthetic pulses (see also Table 
\ref{tabresults}). This is marginally due to the temperature in TPs 
being slightly higher in these models, and mostly due to the effect of metallicity on the 
neutron-capture process itself. The neutron flux is proportional to
$^{22}$Ne/$Z$ and the amount of $^{22}$Ne in low-metallicity AGB stars has a strong   
primary component due to conversion of primary dredged-up $^{12}$C
into $^{14}$N, which is then converted into $^{22}$Ne via
double-$\alpha$ capture in the TPs. Hence the neutron flux increases
with decreasing metallicity because the amount of material
that can capture neutrons (both the light element poisons and the $^{56}$Fe 
seeds) is smaller for lower $Z$. This results in a larger number of
free neutrons and a stronger neutron flux. Also in these cases including TDU in 
the final pulses results in a significant increase in the surface
[Rb/Fe], as well as of [Zr/Fe].

\subsection{Lithium} \label{sec:li}

\begin{figure} \includegraphics[height=.35\textheight,angle=-90]{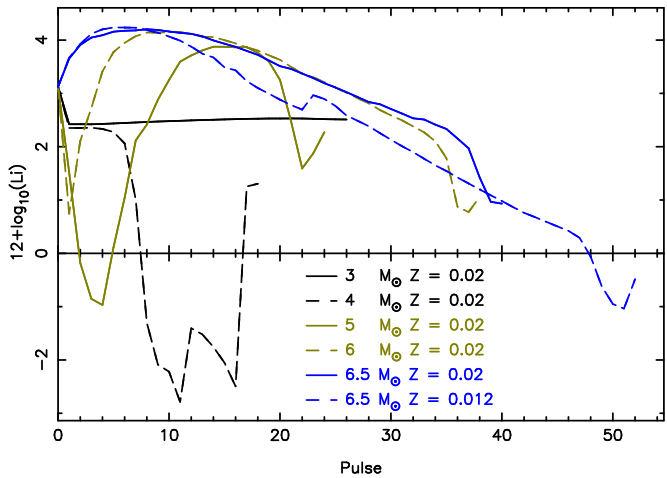}
      \caption{Li abundance, as $\mathrm{log}($Li$/$H$)+12$, versus pulse
number for our solar-metallicity models. (Colour 
figure available online).}
      \label{livspulsenumber}

\end{figure}

The evolution of Li at the stellar surface for our solar metallicity models is shown in 
Fig.~\ref{livspulsenumber}. As the mass of the star increases so does
the temperature at the base of the convective envelope during the AGB
phase (Table~\ref{tab:models}). As the temperature increases the first effect 
on the Li abundance noticeable in the 4, 5, and 6 \msun\ models is the
destruction of Li by HBB. This is because proton captures on $^{7}$Li
are already efficient at temperatures around 20 MK for the low
density conditions of HBB. These temperatures are lower 
than those at which the $^4$He($^3$He,$\gamma$)$^7$Be reaction
is activated (around 40 MK for HBB), which is 
responsible for the production of Li, after electron captures on
$^7$Be. As the temperature at the base of the 
envelope inreases as the star evolves along the AGB, the 
Cameron-Fowler mechanism for the Li production begins,
leading in all the models to a Li-rich phase where the Li 
abundance reaches a maximum of $\mathrm{log}($Li$/$H$)+12 \sim 4$.  
The Li is mixed to hotter temperatures where it is eventually 
destroyed by HBB. In our models we find that the Li-rich phase
lasts for $\approx 10^{5}$ years in a 6 \msun\ model. At some point the 
supply of $^{3}$He runs out and Li production by HBB ceases. 
At this point the star will no longer be observed as Li rich. 
Our results are in agreement with Li production via HBB 
presented by \citet{ventura00} for models of metallicity 0.01. There, 
the Li abundance also first decreases, in the models of mass less than $\sim 6$ \msun, 
and then reaches a maximum close to $\mathrm{log}($Li$/$H$)+12 \sim 4$ before
declining to lower values. From Table~\ref{tabresults} it can be 
noted that for a given initial mass, stars of lower metallicity have a lower 
final Li abundance. This is becasue they reach higher temperature at
the base of the convective envelope (see 
Table~\ref{tab:models}) and destroy Li more efficiently.

\begin{figure} \begin{center}
  \includegraphics[height=.35\textheight,angle=-90]{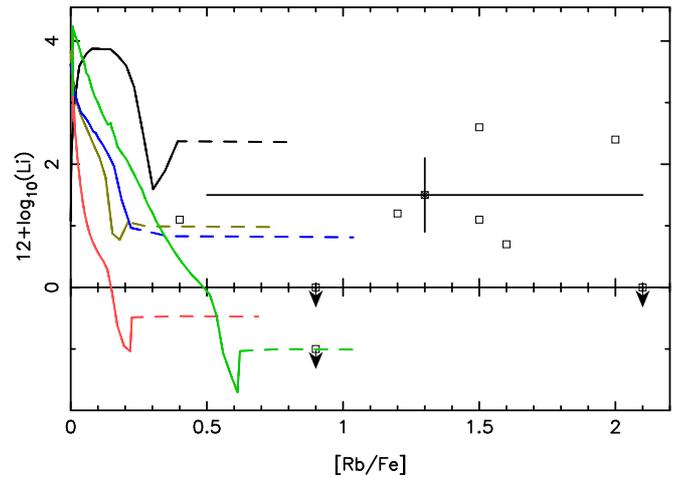}
  \caption{Li abundance versus [Rb/Fe] observed in intermediate-mass
AGB stars (squares) and predicted by our models (solid lines: computed
evolution, dashed lines: synthetic evolution as in Figs.~\ref{rbfehm} and \ref{rbzrhmpmz}). 
Different line colours indicate the following models: 5 \msun\
$Z=0.02$ with PMZ of size 0.0001 \msun\ (black); 6 \msun\ 
$Z=0.02$; (blue) 6.5 \msun\ $Z=0.02$ (dark green); 6.5 \msun\
$Z=0.012$ (red); 6.5 \msun\ $Z=0.012$ with PMZ of 0.0001 \msun\ (light
green). The error bars represent the maximum errors for each
individual data point, though for sake of clarity they are included
for only one data point. Arrows indicate upper
limits. (Colour figure available online).} 
\label{figli} 
\end{center} 
\end{figure}

In Fig.~\ref{figli} we compare predictions for selected models to the observations of stars from 
\citet{garcia06} and \citet{garcia07} for which both Rb and Li data
are available. Shown are also the results for [Rb/Fe] for the
synthetic extension of the models, which as noted above, do not include possible further 
decreases of Li. The observations show large variations in qualitative
agreement with the large changes in Li seen in the models and due to
the strong sensitivity of Li to the temperature of HBB. The stars in the 
Small Magellanic Clouds observed by \citet{plez93} have 
$\mathrm{log}($Li$/$H$)+12$ between $+1.9$ and $+3.5$ and negative [Rb/Fe], which most likely
reflects their initial composition. Our 5 \msun model of $Z=0.02$
would provide a good fit to these observations because HBB is
producing Li and keeping the $^{12}$C/$^{13}$C ratio low, as observed,
but without a significant increase in the [Rb/Fe] abundance. 
However, this metallicity is too high to be compared to the 
Small Magellanic Clouds. For models at the required $Z=0.004$ metallicity 
Rb production occurs already from the lowest masses at which 
HBB is activated (e.g., the 4 \msun\ models at this metallicity 
in Table~\ref{tabresults}). The stars observed by \citet{plez93} 
either (i) are all stars in the early phases of the thermally-pulsing 
AGB where Li production is active but the Rb abundance has yet 
to increase, or (ii) indicate that HBB needs to be more efficient 
and be activated at masses where no substantial Rb production is expected. This may be achieved by 
a different description of convection, either via the Full Spectrum of Turbulence or 
equivalently by using a higher mixing length parameter $\alpha$ \citet{ventura05a}.

\section{Discussion and conclusions}

Our main conclusions can be summarised as follows:

\begin{enumerate}

\item{We predict an increase in the [Rb/Fe] ratio with increasing
stellar mass. This is also suggested by the observations. However, our
standard models are not able to quantitatively  reproduce the high
[Rb/Fe] values found in the observed intermediate-mass OH/IR stars.}

\item{Inclusion of a partial mixing zone (PMZ) increases the final
surface Rb content, but also increases the Zr content, 
resulting in a [Zr/Fe] that is higher than the observed upper limit.}

\item{The uncertainties in the neutron-capture cross sections relevant
to Rb production do not introduce significant enough changes to our 
predicted [Rb/Fe] to match the observed data. Varying the $^{22}$Ne($\alpha,n$)$^{25}$Mg neutron 
source rate produces modest increases to the [Rb/Fe] ratio, but not
enough to match the observed values.}

\item{In order to remove the whole envelope we synthetically
extend our models. By including these final pulses we find [Rb/Fe] 
ratios that are high enough to match the lower end of the observed
data, taking into account the large error bars on the 
data. This relies on the TDU efficiency remaining as high during the
final few TPs when the envelope mass is small as it 
was at the last computed TP. A better understanding of the TDU
efficiency with receding envelope mass is essential and requires 
further investigation. However, one difficulty is that HBB is expected
to be ineffective during this final stage of AGB 
evolution. This means the star will likely become carbon rich if the 
TDU is efficient, while the observed stars are oxygen rich.}

\item{More Rb (in absolute numbers) is produced as the
metallicity decreases. This reproduces the observed 
qualitative trend as derived from recent data for low-metallicity 
intermediate-mass AGB stars in the Magellanic Clouds
\citep{garcia09}. However, also in this case the models cannot 
match the observations quantitatively.}

\end{enumerate}

The main overall difficulty in matching the Rb and Zr observed
abundances is that the $s$ process cannot produce 
differences larger than 0.5 dex between Zr and Rb as these two
elements belong to the same $s$-process peak. On 
the other hand, the observations show differences between these 
elements up to 5 dex. Since the maximum [Rb/Zr] ratio produced 
by the $s$ process is independent of the stellar models and their 
uncertainties, as it derives from the basic operation of the $s$ 
process and the neutron-capture cross sections of the nuclei 
involved (which are know to the level of 5\%), we need to turn to 
observational uncertainties to explain the mismatch for the [Rb/Zr] 
ratio. As discussed in Sect.~\ref{sec:obs} and in \citet{garcia09} 
the solution to the problem may lie with an incomplete 
understanding of the atmospheres of luminous AGB stars. More 
realistic model atmospheres for intermediate-mass O-rich AGB stars 
(e.g., the inclusion of a circumstellar dust envelope and 3D 
hydrodynamical simulations) as well as NLTE calculations 
need to be developed in order to shed some light on the observed 
discrepancy between the observations and the AGB nucleosynthesis 
theoretical predictions. 

Another possibility that needs to be carefully evaluated is whether 
the Rb and Zr abundances in these stars are affected by dust
condensation. The condensation temperature of Zr in a gas of solar-system composition is 1741 K 
\citep{lodders03}, and it would be even higher in a gas where the abundance of Zr is enhanced by the $s$ 
process. It is reasonable to expect that a significant fraction
of the Zr abundance is removed from the gas by condensation into
dust. On the other hand, the condensation temperature of Rb is much lower (800 K), 
and this element is not expected to be removed from the gas. This
effect could explain the observations, which 
samples the gas and not the dust, around the star. A similar idea was put 
forward by \citet{dinerstein06} to explain the non-detection of the 
ground-state fine-structure line of triply ionized Zr in several planetary nebulae 
known to have enhanced abundances of other light neutron-capture
elements. To explain the most extreme observed [Rb/Zr] ratios
requires that most of the Zr and none of the Rb has condensed into dust around these 
stars. This possibility needs to be examined by detailed models of dust formation.

If this ``dust selection'' effect is found to be significant, then
stellar model uncertainties can also be invoked to play a role in
the mismatch between data and predictions with regards to the [Rb/Fe] ratio. For 
example, the inclusion of a PMZ in intermediate-mass AGB stars may
provide a solution to the Rb abundances investigated here, 
although as discussed in Sect.~\ref{sec:models} this would be at odds
with the current modelling of the PMZ in intermediate-mass AGB stars.
Another complementary solution may be found by increasing 
the number of TPs and TDUs episodes experienced by the stellar
models. This number depends on the mass-loss rate of intermediate-mass
AGB stars, which is also highly uncertain. \citet{vw93} noted that there are optically 
bright long period variable stars with periods of $\approx 750$ days 
that are probably intermediate-mass stars of $\sim 5$ \msun. In order
to prevent their mass-loss prescription from removing these observed objects, 
\citet{vw93} recommend a modification to delay the onset of the
superwind in stars of masses greater than 2.5 \msun. Note that in the
models presented here we have not used this modification, which would keep the mass-loss 
rate low for a longer time and would result in more TPs and TDU
episodes. It would also keep the envelope mass relatively high,
allowing for HBB to operate efficiently for more TPs. The outcome would likely be higher levels 
of Rb (and Zr) production in an O-rich envelope. 
Also, Rb production in AGB stars with masses greater than 6.5 \msun\ need to be computed.
Finally, a higher  temperature in the TP would lead to higher Rb production. This may be
related to the penetration of flash-driven convection deeper into the core due to the
inclusion of hydrodynamical overshoot. This would increase the
efficiency of the $^{22}$Ne neutron source as discussed by 
\citet{lugaro03a}. Detailed modelling is needed to explore the viability of these scenarios.

Future observations of the isotopic composition of Mg in the stars
considered here would provide further observational evidence that the
$^{22}$Ne neutron source is activated in these stars as the
neutron-rich isotopes of Mg, $^{25}$Mg and $^{26}$Mg are expected to
be enhanced due to direct production via the $^{22}$Ne$+\alpha$ reactions \citep[][and  
Table~\ref{tabresultslight}]{karakas03b,karakas06a}. Several more elements
could be measured in AGB stars, and N would be particularly
interesting to constrain HBB. Also the derivation of other $s$-process
elements, e.g., Ba, La, Ce, and Pb, is possible and 
could help to further constrain the models. Though the predictions 
presented here involved a nuclear network up to Nb, we are currently
working to extend these models up to Pb. AGB stars have complex 
atmospheres so which elements can actually be measured depends on the
temperature, metallicity, and abundance for each individual star.

Finally, the observations and models discussed here indicate that
intermediate-mass AGB stars synthesize large amounts of 
Rb and that more Rb (in absolute numbers) is produced when 
lowering the metallicity. From this prediction, we extrapolate that 
intermediate-mass AGB model of metallicities typical of Globular
Clusters ($Z \sim 10^{-3}$) will produce even larger abundances of Rb 
than presented here. If intermediate-mass AGB stars produced the heavy elements in Globular 
Cluster stars, then this would rule out these stars as candidates for 
the O, Na, Mg, and Al anomalies. This is because no correlation has
been observed between the heavy elements and the light elements in Globular Cluster stars 
\citep{yong06b,yong08a,yong08b}.  A preliminary study by
\citet{karakas10b} appears to confirm our predictions on the 
basis of two intermediate-mass AGB models, however, a larger set of 
detailed models at the relevant metallicities needs 
to be calculated. Furthermore, we also need to account for the 
data of \citet{plez93}, which indicates that at metallicities
appropriate to the Magellanic Clouds there is a mass or evolutionary 
time when HBB is active but not the $^{22}$Ne neutron source. A dedicated analysis of
these stars will be the focus of a future work.

\begin{acknowledgements}

We thank Carlos Abia and Martin Asplund for discussions. We thank the anonymous 
referee for a very detailed report, which helped to improve the paper
and to broaden the discussion. ML is an ARC 
Future Fellow. AIK is a Stromlo Fellow. DAGH acknowledges support 
provided by the Spanish Ministry of Science and Innovation (MICINN) 
under a JdC grant and under grant AYA-2007-64748.

\end{acknowledgements}

\bibliographystyle{aa} 

\bibliography{library}

\begin{thebibliography}{66}
\expandafter\ifx\csname natexlab\endcsname\relax\def\natexlab#1{#1}\fi

\bibitem[{{Abia} {et~al.}(1993){Abia}, {Boffin}, {Isern}, \& {Rebolo}}]{abia93}
{Abia}, C., {Boffin}, H.~M.~J., {Isern}, J., \& {Rebolo}, R. 1993, \aap, 272,
  455

\bibitem[{{Abia} {et~al.}(2001){Abia}, {Busso}, {Gallino}, {Dom{\'{\i}}nguez},
  {Straniero}, \& {Isern}}]{abia01}
{Abia}, C., {Busso}, M., {Gallino}, R., {et~al.} 2001, \apj, 559, 1117

\bibitem[{{Anders} \& {Grevesse}(1989)}]{anders89}
{Anders}, E. \& {Grevesse}, N. 1989, \gca, 53, 197

\bibitem[{{Angulo} {et~al.}(1999){Angulo}, {Arnould}, {Rayet}, {Descouvemont},
  {Baye}, {Leclercq-Willain}, {Coc}, {Barhoumi}, {Aguer}, {Rolfs}, {Kunz},
  {Hammer}, {Mayer}, {Paradellis}, {Kossionides}, {Chronidou}, {Spyrou},
  {degl'Innocenti}, {Fiorentini}, {Ricci}, {Zavatarelli}, {Providencia},
  {Wolters}, {Soares}, {Grama}, {Rahighi}, {Shotter}, \& {Lamehi
  Rachti}}]{angulo99}
{Angulo}, C., {Arnould}, M., {Rayet}, M., {et~al.} 1999, Nuclear Physics A,
  656, 3

\bibitem[{{Arlandini} {et~al.}(1999){Arlandini}, {K{\"a}ppeler}, {Wisshak},
  {Gallino}, {Lugaro}, {Busso}, \& {Straniero}}]{arlandini99}
{Arlandini}, C., {K{\"a}ppeler}, F., {Wisshak}, K., {et~al.} 1999, \apj, 525,
  886

\bibitem[{{Asplund}(2005)}]{asplund05}
{Asplund}, M. 2005, \araa, 43, 481

\bibitem[{{Bao} {et~al.}(2000){Bao}, {Beer}, {K{\"a}ppeler}, {Voss}, {Wisshak},
  \& {Rauscher}}]{bao00}
{Bao}, Z.~Y., {Beer}, H., {K{\"a}ppeler}, F., {et~al.} 2000, Atomic Data and
  Nuclear Data Tables, 76, 70

\bibitem[{{Beer}(1991)}]{beer91}
{Beer}, H. 1991, \apj, 375, 823

\bibitem[{{Beer} \& {Macklin}(1989)}]{beer89}
{Beer}, H. \& {Macklin}, R.~L. 1989, \apj, 339, 962

\bibitem[{{Boothroyd} {et~al.}(1993){Boothroyd}, {Sackmann}, \&
  {Ahern}}]{boothroyd93}
{Boothroyd}, A.~I., {Sackmann}, I.-J., \& {Ahern}, S.~C. 1993, \apj, 416, 762

\bibitem[{{Busso} {et~al.}(2001){Busso}, {Gallino}, {Lambert}, {Travaglio}, \&
  {Smith}}]{busso01}
{Busso}, M., {Gallino}, R., {Lambert}, D.~L., {Travaglio}, C., \& {Smith},
  V.~V. 2001, \apj, 557, 802

\bibitem[{{Busso} {et~al.}(1999){Busso}, {Gallino}, \& {Wasserburg}}]{busso99}
{Busso}, M., {Gallino}, R., \& {Wasserburg}, G.~J. 1999, ARAA, 37, 239

\bibitem[{{Cameron} \& {Truran}(1977)}]{cameron77}
{Cameron}, A.~G.~W. \& {Truran}, J.~W. 1977, Icarus, 30, 447

\bibitem[{{Cannon}(1993)}]{cannon93}
{Cannon}, R.~C. 1993, \mn, 263, 817

\bibitem[{{Collet} {et~al.}(2011){Collet}, {Hayek}, {Asplund}, {Nordlund},
  {Trampedach}, \& {Gudiksen}}]{collet11}
{Collet}, R., {Hayek}, W., {Asplund}, M., {et~al.} 2011, \aap, 528, A32+

\bibitem[{{Dinerstein} {et~al.}(2006){Dinerstein}, {Lacy}, {Sellgren}, \&
  {Sterling}}]{dinerstein06}
{Dinerstein}, H.~L., {Lacy}, J.~H., {Sellgren}, K., \& {Sterling}, N.~C. 2006,
  in Bulletin of the American Astronomical Society, Vol.~38, American
  Astronomical Society Meeting Abstracts, 156.09--+

\bibitem[{{Federman} {et~al.}(2004){Federman}, {Knauth}, \&
  {Lambert}}]{federman04}
{Federman}, S.~R., {Knauth}, D.~C., \& {Lambert}, D.~L. 2004, \apjl, 603, L105

\bibitem[{{Frost} \& {Lattanzio}(1996)}]{frost96}
{Frost}, C.~A. \& {Lattanzio}, J.~C. 1996, \apj, 473, 383

\bibitem[{{Gallino} {et~al.}(1998){Gallino}, {Arlandini}, {Busso}, {Lugaro},
  {Travaglio}, {Straniero}, {Chieffi}, \& {Limongi}}]{gallino98}
{Gallino}, R., {Arlandini}, C., {Busso}, M., {et~al.} 1998, \apj, 497, 388

\bibitem[{{Garc\'ia-Hern\'andez} {et~al.}(2006){Garc\'ia-Hern\'andez},
  {Garc\'ia-Lario}, {Plez}, {D'Antona}, {Manchado}, \&
  {Trigo-Rodriguez}}]{garcia06}
{Garc\'ia-Hern\'andez}, D.~A., {Garc\'ia-Lario}, P., {Plez}, B., {et~al.} 2006,
  Science, 314, 1751

\bibitem[{{Garc\'ia-Hern\'andez} {et~al.}(2007){Garc\'ia-Hern\'andez},
  {Garc\'ia-Lario}, {Plez}, {Manchado}, {D'Antona}, {Lub}, \&
  {Habing}}]{garcia07}
{Garc\'ia-Hern\'andez}, D.~A., {Garc\'ia-Lario}, P., {Plez}, B., {et~al.} 2007,
  \aap, 462, 711

\bibitem[{{Garc\'ia-Hern\'andez} {et~al.}(2009){Garc\'ia-Hern\'andez},
  {Manchado}, Lambert, Plez, García-Lario, D'Antona, Lugaro, Karakas, \& van
  Raai}]{garcia09}
{Garc\'ia-Hern\'andez}, D.~A., {Manchado}, A., Lambert, D.~L., {et~al.} 2009,
  \apj, 705, L31

\bibitem[{{Goriely}(1999)}]{goriely99}
{Goriely}, S. 1999, \aap, 342, 881

\bibitem[{{Goriely} \& {Mowlavi}(2000)}]{goriely00}
{Goriely}, S. \& {Mowlavi}, N. 2000, \aap, 362, 599

\bibitem[{{Heil} {et~al.}(2008){Heil}, {K{\"a}ppeler}, {Uberseder}, {Gallino},
  {Bisterzo}, \& {Pignatari}}]{heil08}
{Heil}, M., {K{\"a}ppeler}, F., {Uberseder}, E., {et~al.} 2008, \prc, 78,
  025802

\bibitem[{{Herwig}(2004)}]{herwig04a}
{Herwig}, F. 2004, \apj, 605, 425

\bibitem[{{Herwig}(2005)}]{herwig05}
{Herwig}, F. 2005, ARAA, 43, 435

\bibitem[{{Hollowell} {et~al.}(1990){Hollowell}, {Iben}, \&
  {Fujimoto}}]{hollowell90}
{Hollowell}, D., {Iben}, I.~J., \& {Fujimoto}, M.~Y. 1990, \apj, 351, 245

\bibitem[{{Iben}(1975)}]{iben75a}
{Iben}, Jr., I. 1975, \apj, 196, 525

\bibitem[{{Iben} \& {Renzini}(1983)}]{iben+renzini83}
{Iben}, I., J. \& {Renzini}, A. 1983, ARAA, 21, 271

\bibitem[{{Iliadis} {et~al.}(2010){Iliadis}, {Longland}, {Champagne}, {Coc}, \&
  {Fitzgerald}}]{iliadis10}
{Iliadis}, C., {Longland}, R., {Champagne}, A.~E., {Coc}, A., \& {Fitzgerald},
  R. 2010, Nuclear Physics A, 841, 31

\bibitem[{{K\"{a}ppeler} {et~al.}(1989){K\"{a}ppeler}, {Beer}, \&
  {Wisshak}}]{kaeppeler89}
{K\"{a}ppeler}, F., {Beer}, H., \& {Wisshak}, K. 1989, Reports on Progress in
  Physics, 52, 945

\bibitem[{{Karakas} \& {Lattanzio}(2007)}]{karakas07b}
{Karakas}, A. \& {Lattanzio}, J.~C. 2007, PASA, 24, 103

\bibitem[{{Karakas}(2010)}]{karakas10a}
{Karakas}, A.~I. 2010, \mnras, 403, 1413

\bibitem[{{Karakas} {et~al.}(2010){Karakas}, {Campbell}, {Lugaro}, {Yong}, \&
  {Chieffi}}]{karakas10b}
{Karakas}, A.~I., {Campbell}, S.~W., {Lugaro}, M., {Yong}, D., \& {Chieffi}, A.
  2010, \memsai, 81, 1010

\bibitem[{{Karakas} \& {Lattanzio}(2003)}]{karakas03b}
{Karakas}, A.~I. \& {Lattanzio}, J.~C. 2003, \pasa, 20, 279

\bibitem[{{Karakas} {et~al.}(2006){Karakas}, {Lugaro}, {Wiescher}, {Goerres},
  \& {Ugalde}}]{karakas06a}
{Karakas}, A.~I., {Lugaro}, M., {Wiescher}, M., {Goerres}, J., \& {Ugalde}, C.
  2006, \apj, 643, 471

\bibitem[{{Karakas} {et~al.}(2009){Karakas}, {van Raai}, {Lugaro}, {Sterling},
  \& {Dinerstein}}]{karakas09}
{Karakas}, A.~I., {van Raai}, M.~A., {Lugaro}, M., {Sterling}, N.~C., \&
  {Dinerstein}, H.~L. 2009, \apj, 690, 1130

\bibitem[{{Lambert} \& {Luck}(1976)}]{lambert76}
{Lambert}, D.~L. \& {Luck}, R.~E. 1976, The Observatory, 96, 100

\bibitem[{{Lambert} {et~al.}(1995){Lambert}, {Smith}, {Busso}, {Gallino}, \&
  {Straniero}}]{lambert95}
{Lambert}, D.~L., {Smith}, V.~V., {Busso}, M., {Gallino}, R., \& {Straniero},
  O. 1995, \apj, 450, 302

\bibitem[{{Lattanzio}(1986)}]{lattanzio86}
{Lattanzio}, J.~C. 1986, \apj, 311, 708

\bibitem[{{Lind} {et~al.}(2011){Lind}, {Asplund}, {Barklem}, \&
  {Belyaev}}]{lind11}
{Lind}, K., {Asplund}, M., {Barklem}, P.~S., \& {Belyaev}, A.~K. 2011, \aap,
  528, A103+

\bibitem[{{Lodders}(2003)}]{lodders03}
{Lodders}, K. 2003, \apj, 591, 1220

\bibitem[{{Lugaro} \& {Chieffi}(2011)}]{lugaro11}
{Lugaro}, M. \& {Chieffi}, A. 2011, in Lecture Notes in Physics, ed. {R.~Diehl,
  D.~H.~Hartmann, \& N.~Prantzos}, Vol. 812 (Berlin Springer Verlag), 83--152

\bibitem[{{Lugaro} {et~al.}(2003){Lugaro}, {Herwig}, {Lattanzio}, {Gallino}, \&
  {Straniero}}]{lugaro03a}
{Lugaro}, M., {Herwig}, F., {Lattanzio}, J.~C., {Gallino}, R., \& {Straniero},
  O. 2003, \apj, 586, 1305

\bibitem[{{Lugaro} {et~al.}(2007){Lugaro}, {Karakas}, {Nittler}, {Alexander},
  {Hoppe}, {Iliadis}, \& {Lattanzio}}]{lugaro07}
{Lugaro}, M., {Karakas}, A.~I., {Nittler}, L.~R., {et~al.} 2007, \aap, 461, 657

\bibitem[{{Lugaro} {et~al.}(2004){Lugaro}, {Ugalde}, {Karakas}, {G{\"o}rres},
  {Wiescher}, {Lattanzio}, \& {Cannon}}]{lugaro04}
{Lugaro}, M., {Ugalde}, C., {Karakas}, A.~I., {et~al.} 2004, \apj, 615, 934

\bibitem[{{Meyer}(1994)}]{meyer94}
{Meyer}, B.~S. 1994, ARAA, 32, 153

\bibitem[{{Plez} {et~al.}(1993){Plez}, {Smith}, \& {Lambert}}]{plez93}
{Plez}, B., {Smith}, V.~V., \& {Lambert}, D.~L. 1993, \apj, 418, 812

\bibitem[{{Reimers}(1975)}]{reimers75}
{Reimers}, D. 1975, {Circumstellar envelopes and mass loss of red giant stars}
  (Springer-Verlag New York), 229--256

\bibitem[{{Siess} {et~al.}(2004){Siess}, {Goriely}, \& {Langer}}]{siess04}
{Siess}, L., {Goriely}, S., \& {Langer}, N. 2004, \aap, 415, 1089

\bibitem[{{Simmerer} {et~al.}(2004){Simmerer}, {Sneden}, {Cowan}, {Collier},
  {Woolf}, \& {Lawler}}]{simmerer04}
{Simmerer}, J., {Sneden}, C., {Cowan}, J.~J., {et~al.} 2004, \apj, 617, 1091

\bibitem[{{Smith} \& {Lambert}(1990)}]{smith90b}
{Smith}, V.~V. \& {Lambert}, D.~L. 1990, \apj, 361, L69

\bibitem[{{Stancliffe} \& {Jeffery}(2007)}]{stancliffe07b}
{Stancliffe}, R.~J. \& {Jeffery}, C.~S. 2007, \mn, 375, 1280

\bibitem[{{Straniero} {et~al.}(1997){Straniero}, {Chieffi}, {Limongi}, {Busso},
  {Gallino}, \& {Arlandini}}]{straniero97}
{Straniero}, O., {Chieffi}, A., {Limongi}, M., {et~al.} 1997, \apj, 478, 332

\bibitem[{{Straniero} {et~al.}(2000){Straniero}, {Limongi}, {Chieffi},
  {Dominguez}, {Busso}, \& {Gallino}}]{straniero00}
{Straniero}, O., {Limongi}, M., {Chieffi}, A., {et~al.} 2000, \memsai, 71, 719

\bibitem[{{Thielemann} {et~al.}(1986){Thielemann}, {Truran}, \&
  {Arnould}}]{thielemann86}
{Thielemann}, F.-K., {Truran}, J.~W., \& {Arnould}, M. 1986, in Advances in
  Nuclear Astrophysics, ed. E.~{Vangioni-Flam}, J.~{Audouze}, M.~{Casse}, J.-P.
  {Chieze}, \& J.~{Tran Thanh van} (Gif-sur-Yvette, France, Editions
  Frontieres), 525--540

\bibitem[{{Travaglio} {et~al.}(2004){Travaglio}, {Gallino}, {Arnone}, {Cowan},
  {Jordan}, \& {Sneden}}]{travaglio04}
{Travaglio}, C., {Gallino}, R., {Arnone}, E., {et~al.} 2004, \apj, 601, 864

\bibitem[{{Vassiliadis} \& {Wood}(1993)}]{vw93}
{Vassiliadis}, E. \& {Wood}, P.~R. 1993, \apj, 413, 641

\bibitem[{{Ventura} \& {D'Antona}(2005{\natexlab{a}})}]{ventura05a}
{Ventura}, P. \& {D'Antona}, F. 2005{\natexlab{a}}, \aa, 431, 279

\bibitem[{{Ventura} \& {D'Antona}(2005{\natexlab{b}})}]{ventura05b}
{Ventura}, P. \& {D'Antona}, F. 2005{\natexlab{b}}, \aa, 439, 1075

\bibitem[{{Ventura} {et~al.}(2000){Ventura}, {D'Antona}, \&
  {Mazzitelli}}]{ventura00}
{Ventura}, P., {D'Antona}, F., \& {Mazzitelli}, I. 2000, \aap, 363, 605

\bibitem[{{Wood} {et~al.}(1983){Wood}, {Bessell}, \& {Fox}}]{wood83}
{Wood}, P.~R., {Bessell}, M.~S., \& {Fox}, M.~W. 1983, \apj, 272, 99

\bibitem[{{Yong} {et~al.}(2006){Yong}, {Aoki}, {Lambert}, \&
  {Paulson}}]{yong06b}
{Yong}, D., {Aoki}, W., {Lambert}, D.~L., \& {Paulson}, D.~B. 2006, \apj, 639,
  918

\bibitem[{{Yong} {et~al.}(2008{\natexlab{a}}){Yong}, {Karakas}, {Lambert},
  {Chieffi}, \& {Limongi}}]{yong08b}
{Yong}, D., {Karakas}, A.~I., {Lambert}, D.~L., {Chieffi}, A., \& {Limongi}, M.
  2008{\natexlab{a}}, \apj, 689, 1031

\bibitem[{{Yong} {et~al.}(2008{\natexlab{b}}){Yong}, {Lambert}, {Paulson}, \&
  {Carney}}]{yong08a}
{Yong}, D., {Lambert}, D.~L., {Paulson}, D.~B., \& {Carney}, B.~W.
  2008{\natexlab{b}}, \apj, 673, 854

\end{thebibliography}

\end{document}